\documentclass[
reprint,
superscriptaddress,
showpacs,
 amsmath,amssymb,
 aps,
prl,
floatfix,
]{revtex4-1}
\usepackage{color}
\usepackage{graphicx}% Include figure files
\usepackage{dcolumn}% Align table columns on decimal point
\usepackage{bm}% bold math
\usepackage{hyperref}% add hypertext capabilities
\usepackage{soul}
\begin{document}

\preprint{APS/123-QED}

\title{Directional bonding explains high conductance values of atomic contacts in bcc metals}

\author{W. Dednam}
\email{wd2@alu.ua.es}
\affiliation{Department of Physics, Science Campus, University of South Africa, Private Bag X6, Florida Park 1710, South Africa}
\affiliation{Departamento de F\'\i sica Aplicada and Unidad asociada CSIC, Universidad de Alicante, Campus de San Vicente del Raspeig, E-03690 Alicante, Spain}

\author{C. Sabater}
\email{carlos.sabater@ua.es}
\affiliation{Departamento de F\'\i sica Aplicada and Unidad asociada CSIC, Universidad de Alicante, Campus de San Vicente del Raspeig, E-03690 Alicante, Spain}

\author{M. R. Calvo}
\affiliation{Departamento de F\'\i sica Aplicada and Unidad asociada CSIC, Universidad de Alicante, Campus de San Vicente del Raspeig, E-03690 Alicante, Spain}

\author{C. Untiedt}
\affiliation{Departamento de F\'\i sica Aplicada and Unidad asociada CSIC, Universidad de Alicante, Campus de San Vicente del Raspeig, E-03690 Alicante, Spain}
 
\author{J. J. Palacios}
\affiliation{Departamento de F\'\i sica de la Materia Condensada, Condensed Matter Physics Center (IFIMAC), and
Instituto Nicol\'as Cabrera,  Universidad
Aut\'onoma de Madrid, 28049 Madrid, Spain} 
 
\author{A. E. Botha}
\affiliation{Department of Physics, Science Campus, University of South Africa, Private Bag X6, Florida Park 1710, South Africa}
 
\author{M. J. Caturla}
\affiliation{Departamento de F\'\i sica Aplicada and Unidad asociada CSIC, Universidad de Alicante, Campus de San Vicente del Raspeig, E-03690 Alicante, Spain}

\date{\today}

\begin{abstract}
Atomic-sized junctions of iron, created by controlled rupture, present unusually high values of conductance compared to other metals. This result is counter-intuitive since, at the nanoscale, body- centered cubic metals are expected to exhibit lower coordination than face-centered cubic metals. In this work, classical molecular dynamics simulations of contact rupture, using an interatomic potential that accounts for directional bonding, yield highly-coordinated stable structures before rupture, unlike an isotropic bonding potential, which results in the expected stable single-atom contacts. Density functional theory electronic transport calculations show that conductance values of these highly coordinated and highly stable structures, can explain the experimentally measured values for conductance of body-centered cubic atomic contacts, thus revealing the important role of directional bonding in these metals.
\end{abstract}

\maketitle

Stretching a metallic nanowire results in a progressive reduction of its cross-section at the weakest point, until it finally breaks. From an atomistic viewpoint, when the minimum cross-section of the nanowire contains only a few atoms, and for very slow stretching, the minimum cross-section can in fact decrease by one atom at a time  \cite{Landman1990,Agrait2003}.
It seems reasonable to assume that the ultimate \textit{stable} contact that holds the metal together is a single atom. Indeed, measurements of conductance for atomically sharp contacts seem to point in this direction since the stable contact conductance before rupture for most metals is just above one quantum, except for some notable exceptions; such as, iron \cite{Otal13,CalvoNature}, tantalum, molybdenum and tungsten \cite{DenBoer2007,Halbritter2003}. In fact, it is still widely assumed that the chemical valence of the bridging atom in single-atom contacts primarily determines the pre-rupture conductance values \cite{Requist2016,Scheer1998}.

Atomic-sized contacts are typically realized via mechanically controllable break junctions (MCBJ) \cite{MCBJ1,MCBJ2}  or scanning tunneling microscope break junctions (STM-BJ) \cite{NachoSTMatom93,AgraitSTM93} and characterized through electron transport  measurements. Classical molecular dynamics (CMD) simulations and first principles transport calculations have been key in providing interpretations of the experimental results \cite{Landman1990}. In the past decade, through the combination of experiments and simulations, researchers have identified different atomic contact geometries that may form just before rupture \cite{Serena2005,Serena2001,Mochales2005, Understanding2013} and stated the important role played by the nearest neighbor atoms in the mechanical and electrical properties of these structures \cite{SAbaterRole2018}.

Computationally, most of the systems studied so far have been face-centered cubic (FCC) crystals, which exhibit a high probability to form single-atom contacts immediately before rupture. However, recent calculations \cite{FeOTal2016} show a clear mismatch between experiment and theory in the case of body-centered cubic (BCC) iron. The reason behind this discrepancy is not clear, since three factors could play a role: chemical valence, crystal lattice structure and/or magnetism.

To shed light on this question, we first perform CMD simulations of the iron rupture process, using two different interatomic potentials: one in which the bonding between atoms is treated as isotropic, the other in which the bonds also have covalent character. Density functional theory (DFT) calculations on CMD snapshots of the atomic configurations are then used to obtain the electronic transport properties (the conductance) and to compare with experimental data obtained from electron transport experiments in an STM-BJ at low temperature (4.2 K).

\begin{figure}[htp]
\centering
\includegraphics[width=0.48\textwidth]{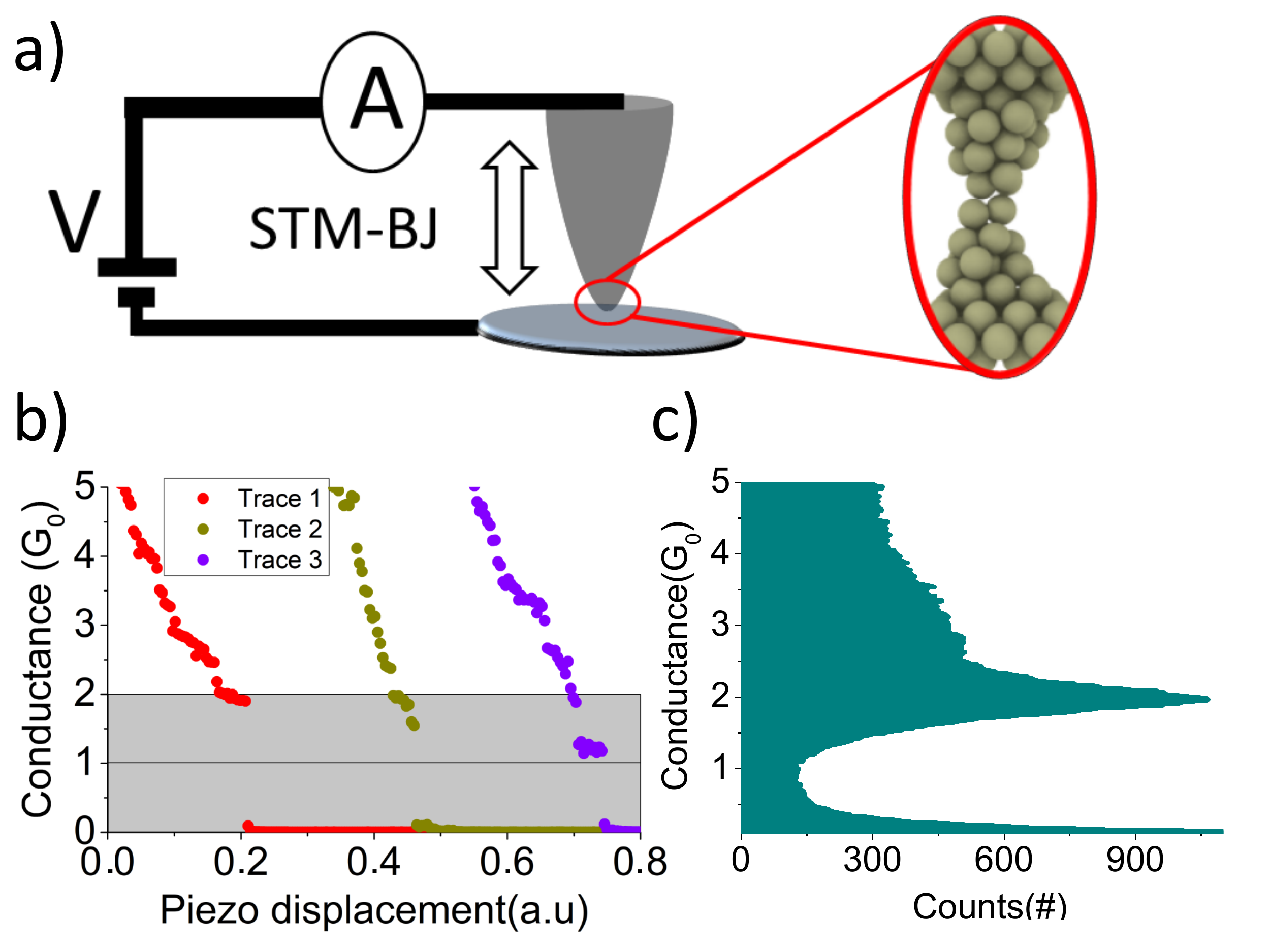}
\caption{a) Experimental STM-BJ setup b) Traces of conductance c) Histogram of conductance constructed from more than five hundred rupture traces of iron.}
\label{fig:Expsetup}
\end{figure}

In the STM-BJ configuration, the electrode tip can collide with and be withdrawn from the surface over continuous cycles of rupture and formation of the atomic-sized contact. 
The electrode tip consists of iron wire of nominal diameter 0.25 mm (with a purity of 99.99\%), and is connected in an electrical circuit as shown in Fig. \ref{fig:Expsetup} a), where a constant bias voltage $V=100$ mV is connected in series with the IV converter amplifier and, in turn, to a resistor of interest (in our case an atomic-sized contact). We express conductance that is the inverse of the resistance, in units of the quantum of conductance $G_0 = 2e^2/h$, where $e$ is the charge of the electron,
$h$ is the Planck constant and the factor of 2 accounts for the spin degeneracy.

Typically, we record the conductance  as a function of the relative displacement between the electrodes. The resulting curves are called rupture traces, as shown in Fig. \ref{fig:Expsetup} b). From each trace, we can build a histogram of conductance, which reveals the most frequent conductance values that the trace contains. Upon accumulating a significant number of individual histograms (one for every trace), we can construct a full histogram of conductance. For iron, it exhibits a clear peak at $\approx 2G_0$ (see, Fig. \ref{fig:Expsetup} c)), as has been previously observed \cite{UntiedtFECONIMAG,FeOTal2016,CalvoNature}.

In CMD simulations, the trajectory of each atom in the contact can be obtained by solving Newton's second law by using a suitable interatomic potential to describe the interactions between the atoms. This constitutes the basic principle of classical molecular dynamics \cite{ART_MDsim2004}, which provides us with the means to model a very wide variety of materials down to the atomic level.  The real problem here, however, is to select the most appropriate semi-empirical interatomic potential \cite{daw1983semiempirical}, such that the interactions between the iron atoms are described with as much detail as possible.
By far the most extensively used empirical potential to model metals is the embedded-atom method (EAM) potential \cite{Malerba2010,SAbaterRole2018,Understanding2013,FeOTal2016,CuevasBook2010,Agrait2003}, in which bonding is assumed to be isotropic. While this assumption is valid for FCC structures, BCC structures have a lower coordination about an individual atom and thus exhibit more directionality in their bonding, i.e. slightly covalent character \cite{pettiford}. Accordingly, the \textit{modified} embedded-atom method (MEAM) potential \cite{Baskes1992} may provide a more realistic description of the bonding in BCC metals, since directionality is included. In this work we thus compare the MEAM and EAM methods in order to determine which will lead to better agreement with experimental results in electronic transport calculations on CMD-generated structures.

\begin{figure}[htp]
\centering
\includegraphics[width=0.48\textwidth]{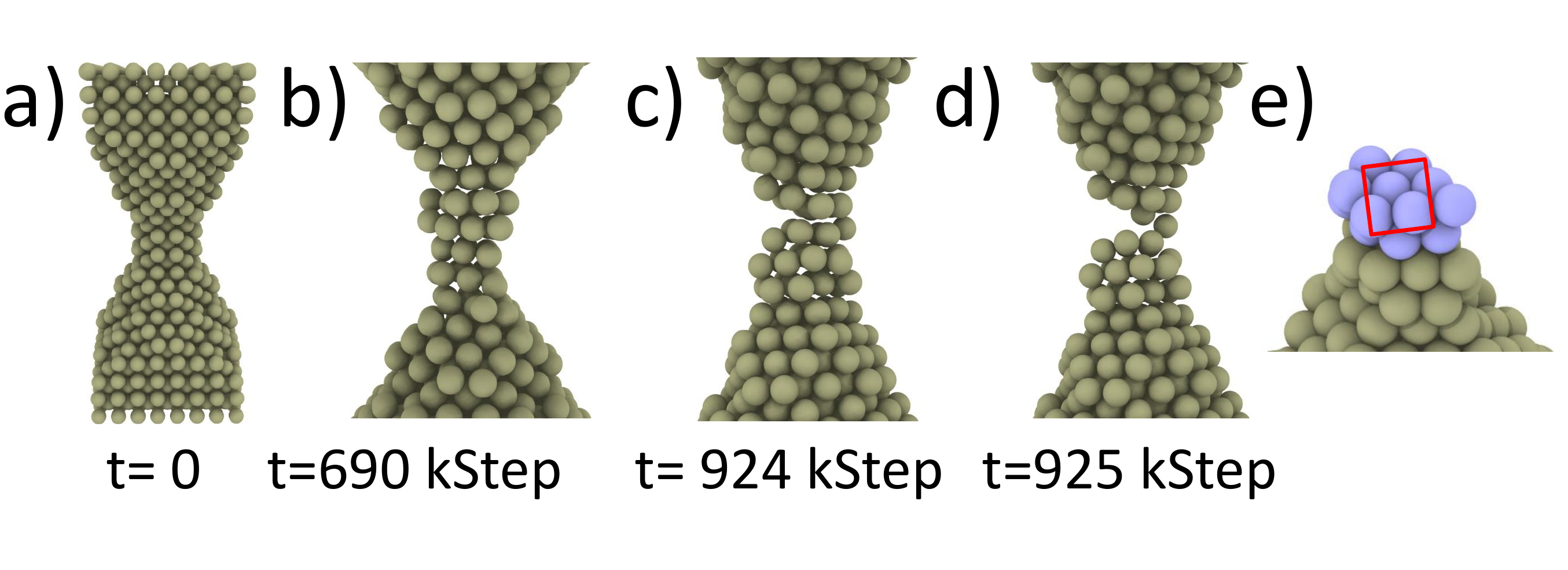}
\caption{a) A typical initial input structure used in the simulations. b) through d) show the process of rupture which occurs in 20 out of the 100 simulations of rupture with the MEAM potential \cite{ETESAMI2018}. The BCC iron contact goes through a crystallographic re-orientation under tension, from having (001) to (110) planes perpendicular to the length of the contact (shown in b)). Rupture occurs in these cases via cleavage of (110)-oriented planes (shown in e)). The whole process lasts only a few picoseconds, e.g., from the structures in c) to d). e) A cutaway (the top half of the contact has been removed) showing the characteristic 5-atom structure (red rectangle) in a (110) surface of a BCC lattice.}
\label{GeFEO}
\end{figure}

For both potentials, we use the Large-scale Atomic/Molecular Massively Parallel Simulator (LAMMPS) \cite{plimpton1995fast,lammps2}. Additionally, to imitate the experimental conditions, all the simulations are realized using a Nos\'e-Hoover thermostat \cite{nose1984molecular,hoover1985canonical} to maintain a constant temperature. The thermostat is applied at the recommended interval of 1000 simulation time steps~\cite{lammps2}. We use a time step of 1 fs and the same initial input structure, consisting of $\approx 1500$ atoms, for comparison of the two potentials. Figure~\ref{GeFEO} shows a representative example of the rupture process of iron using the MEAM potential, with the atoms initially occupying positions in a perfect BCC lattice oriented along the (001) crystallographic direction (see Fig. \ref{GeFEO} a)). The initial velocities of the atoms are randomized at the beginning of each rupture run and correspond to an average temperature of 4.2 K. The input structure is stretched at $\approx 1$ m/s until rupture.  During every single rupture simulation, out of an ensemble of 100 independent runs performed with each potential, we compute the number of atoms in the minimum cross-section of the model contact by means of the Bratkovksy algorithm \cite{Bratkovsky}. The minimum cross-section and simulation trajectory are both recorded every picosecond. For the purpose of comparison, traces are truncated 100 ps before the moment of rupture when constructing cross-section histograms.

\begin{figure}[htp]
\centering
\includegraphics[width=0.48\textwidth]{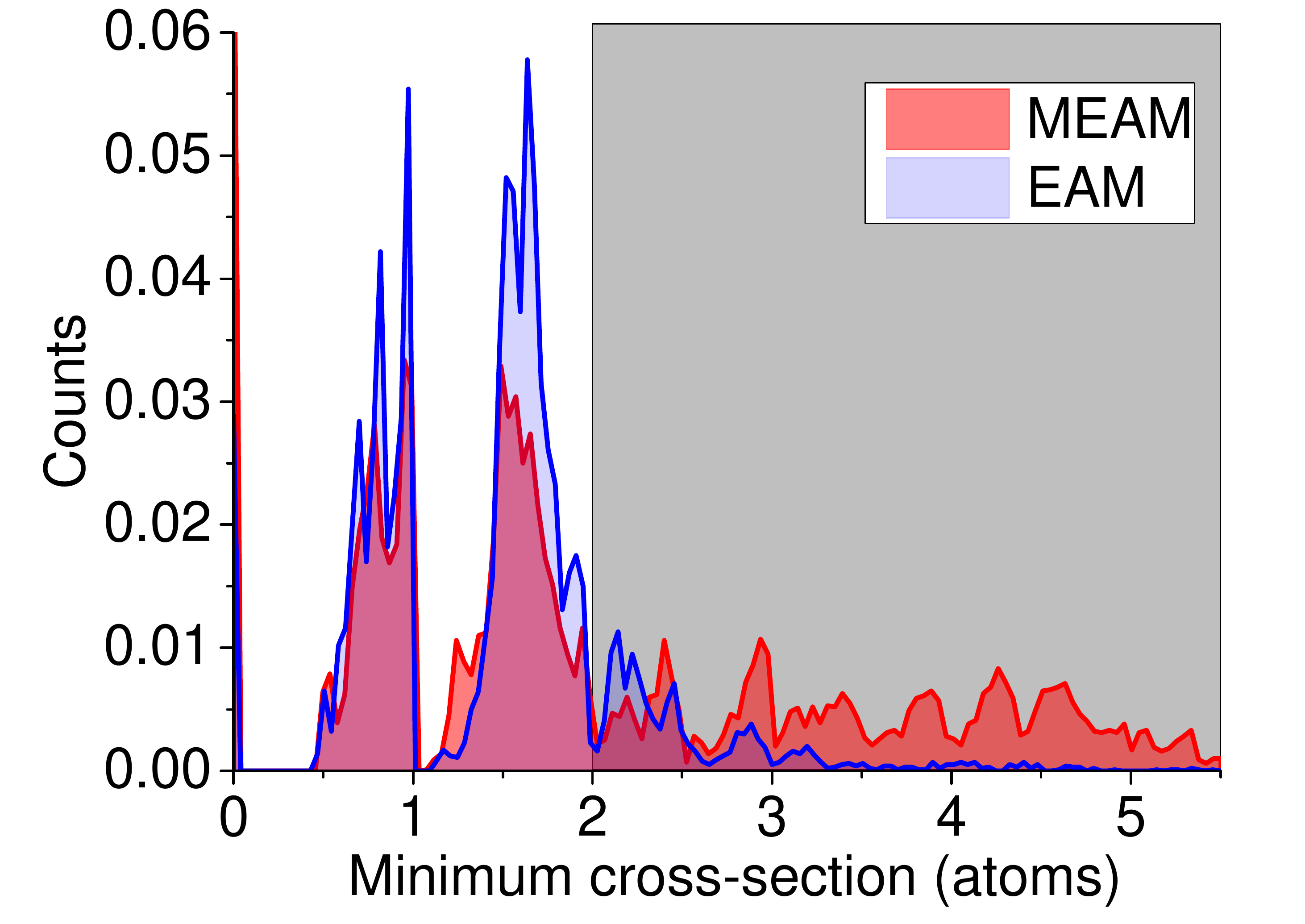}
\caption{Minimum cross-section histograms obtained after 100 rupture simulations using the  MEAM potential \cite{ETESAMI2018}  (pink shading with red outline) and the EAM potential \cite{Malerba2010} (grey shading with blue outline).}
\label{Bratkovsky}
\end{figure}

Figure \ref{Bratkovsky} compares the two normalized histograms of minimum cross-section data obtained with each of the two potentials. The pink-shaded histogram in Fig. \ref{Bratkovsky} has been constructed by using the most recent MEAM interatomic potential, fitted to the melting point of Fe as well as its near-melting point elastic constants \cite{ETESAMI2018}. This particular potential is suitable for simulations of Fe contact rupture because the (001), (110) and (111) exposed surface energies agrees very well with experiments \cite{ETESAMI2018}. For comparison,
see the blue-shaded minimum cross-section histogram in Fig. \ref{Bratkovsky}, obtained using an EAM potential whose surface energies also agree reasonably well with experiment and DFT calculations \cite{Malerba2010}.  Although both potentials show the formation of one and two-atom contacts (first two peaks in the histogram), the MEAM potential produces stable pre-rupture structures with a higher number of atoms in the minimum cross-section than the EAM potential. This is shown in the shaded tail of the histogram, where the probability of forming structures with more than two atoms right before rupture is lower with the EAM potential than with MEAM.

Note that in the case of FCC metals \cite{Serena2001,Serena2005,Mochales2005,SAbaterRole2018}, and even in the structures obtained with the EAM potential for Fe, the simulated cross-section narrows atom by atom during the rupture process. In contrast, the cross-section obtained using the MEAM potential breaks through cleavage across (110) oriented planes as shown in Fig. \ref{GeFEO} e). These (110) planes are formed after a re-orientation of the contact during the applied tension.
 
To obtain the conductance of snapshots extracted from CMD simulations, such as in Fig. \ref{GeFEO} c), we use the electronic transport code Alicante nanotransport (\texttt{ANT.Gaussian}) \cite{palacios2001fullerene,palacios2002transport,louis2003keldysh,ANTG,GAUSSIAN09}. (For more details of the DFT calculations, see the Supplemental Material \cite{SM}.) Conductance values near the peak of the experimental histogram in Fig. \ref{fig:Expsetup} c) are obtained for structures which correspond to those shown in Fig. \ref{GeFEO} c) (see cases marked with an asterisk (*) in table SI of the Sup. Mat. \cite{SM}), that is, those predicted by the MEAM potential. Note that the EAM potential only reproduces the rupture process illustrated in Fig. \ref{GeFEO} in 3 out of the 100 rupture simulations versus 20 out of a 100 in the case of the MEAM potential. 

More revealing than the statistical study presented above, are the so-called Fano factors \cite{CuevasBook2010}. In experiments at low temperatures, low bias voltage and low frequency range (maximum of 400 kHz), the Fano factor $F$ provides a measure of noise suppression relative to the maximum Poissonian value of $2eI$ \cite{CuevasBook2010}. Therefore, the shot noise from ballistic transport of an electron through an atomic-sized contact, is given by $S_{I}=2eIF$, where $I$ is
the bias current and $e$ the electron charge.
On the other hand, in DFT quantum transport calculations, the spin-polarized conductance can be expressed as \cite{CuevasBook2010}:
$G=\frac{G_{0}}{2}\Sigma_{n,\sigma} T_{n,\sigma}$, where $G_{0}$ is the usual spin-degenerate quantum of conductance and $T_{n,\sigma}$ are the individual spin-resolved eigenchannel transmissions \cite{Jacob2006}. Since not only the geometry but also the number of atoms in the constriction of a contact determines
the overall conductance through their valence orbitals, the individual spin-resolved transmission channels can convey information about the atomic structure of the contacts through the Fano factor (a measure of the number of partially open transmission channels in an atomic-sized contact):
\begin{equation}
F=\frac{\Sigma_{n,\sigma} T_{n,\sigma}(1-T_{n,\sigma})}{\Sigma_{n,\sigma} T_{n,\sigma}}
\end{equation}

\begin{figure}[htp]
 \centering
 \includegraphics[width=0.48\textwidth]{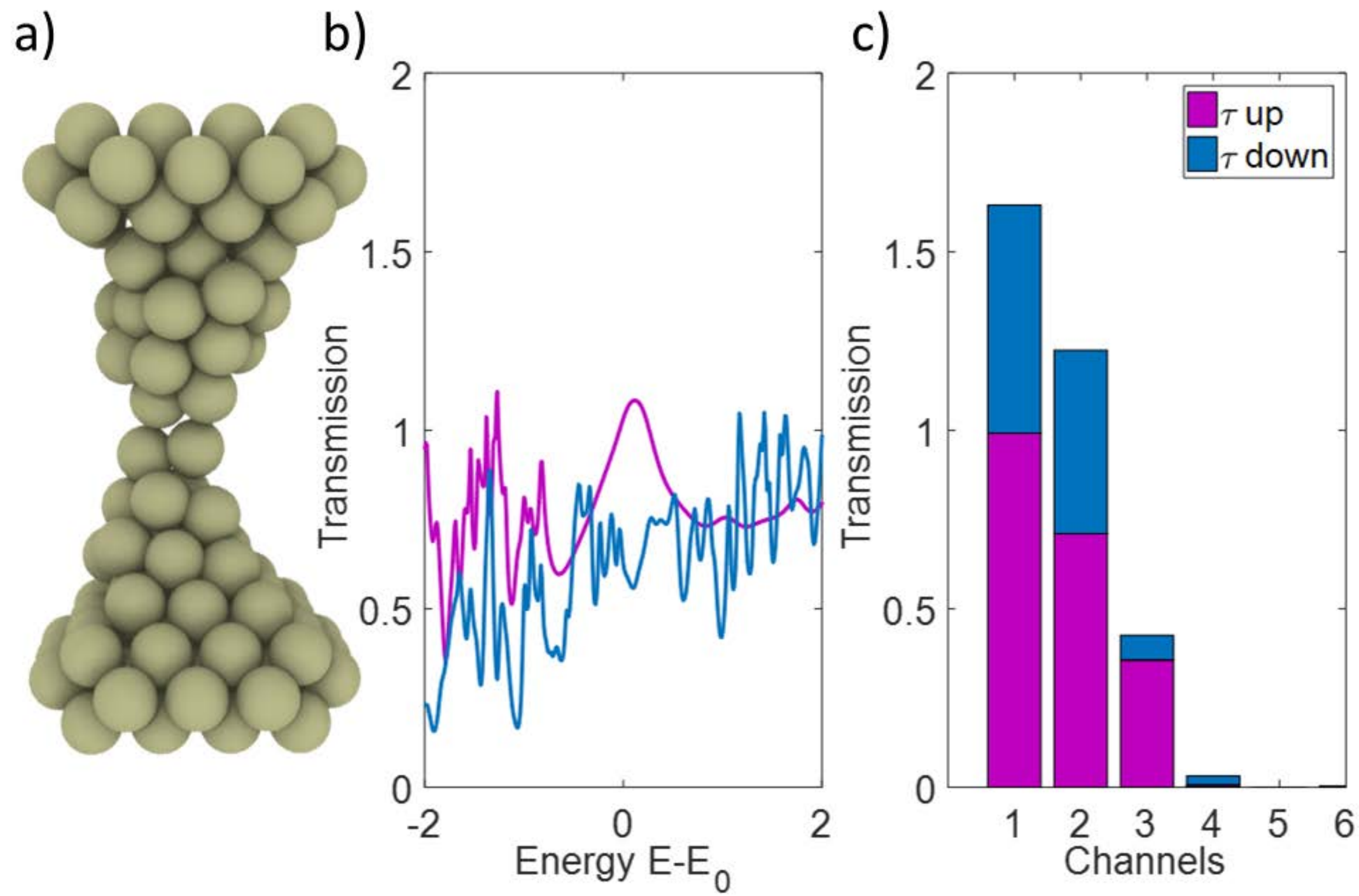} 
 \caption{a) Fe double contact, b) overall spin-resolved transmission vs energy (eV), c) transmission  versus spin eigenchannel. A Fano factor of 0.3084 is obtained from the eigenchannels shown in c).}
 \label{FANOEXAMPLE}
 \end{figure}

Figure \ref{FANOEXAMPLE} shows an example of how an eigenchannel analysis can be carried out to obtain the Fano factor from a conductance calculation on a model contact. A conductance calculation on the structure shown in Fig. \ref{FANOEXAMPLE} a) not only yields the overall spin-resolved transmissions in b) (which sum to $G=1.6574G_0$), but also the contributing spin-resolved eigenchannel transmissions, shown in Fig. \ref{FANOEXAMPLE} c). Therefore, in this case only 5 spin-resolved eigenchannels --3 spin-majority (purple) and 2 spin-minority (blue)-- contribute significantly to the overall transmission. Based on this analysis, one can conclude that at least 2 atoms effectively contribute to the transmission in this contact. The noise in the transmission functions in Fig. \ref{FANOEXAMPLE} b) result from the disorder in CMD structures in general \cite{NiReyes2008} and, in particular, from the \textit{spd} hybridization of the spin-minority channels \cite{Garcia2000iron}.

\begin{figure}[htp]
\centering
\includegraphics[width=0.49\textwidth]{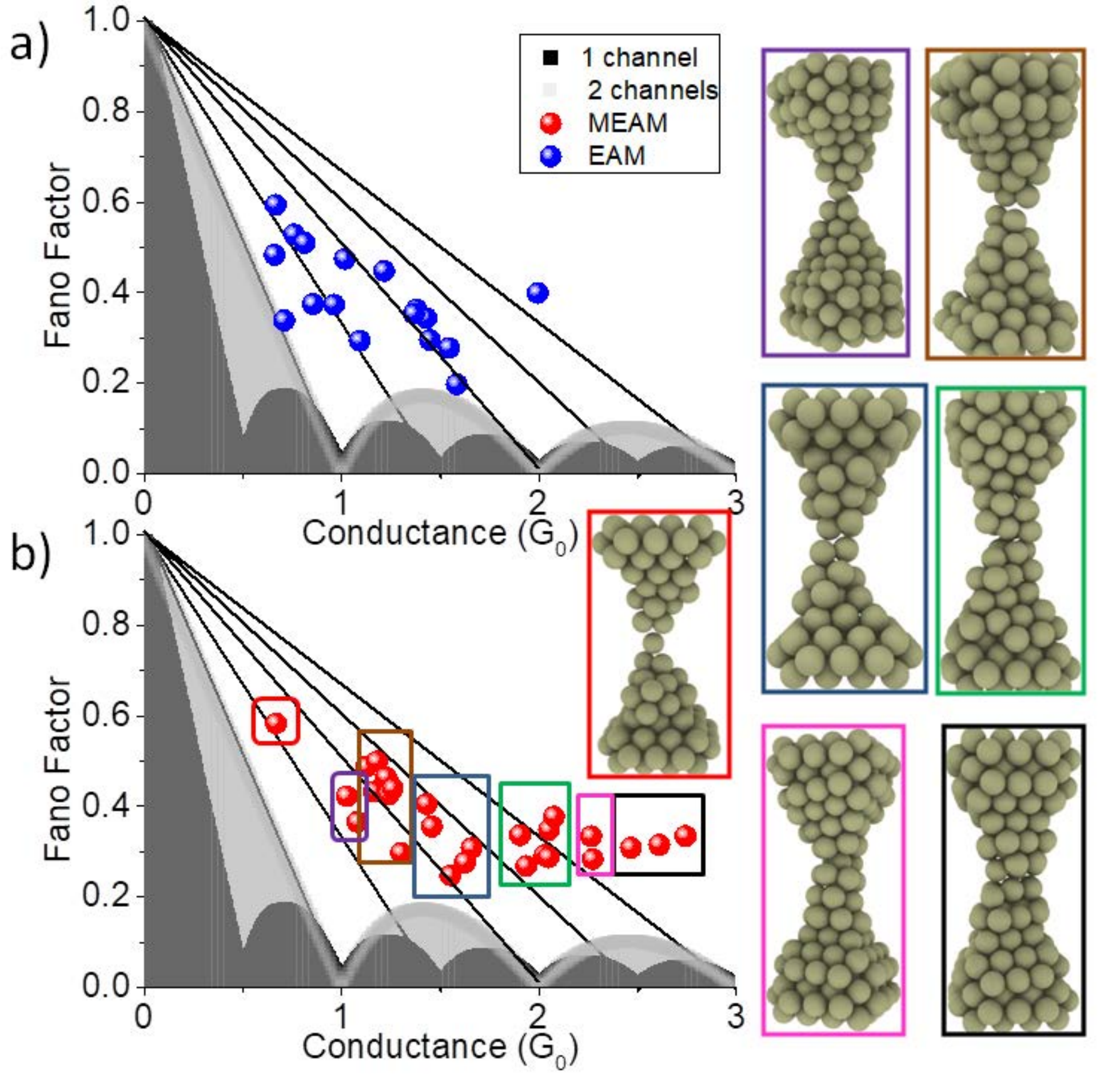}
\caption{Theoretical Fano factors vs calculated conductance for structures obtained via the a) EAM (blue markers) and b) MEAM (red markers) potentials. In the case of the MEAM potential, the data is grouped by color-coded frames according to stable structures just before rupture, shown in the insets. The conductance values are recorded in tables SI and SII \cite{SM}. The diagonal lines determine the number of spin-resolved transmission channels. The agreement with the experimental values in Fig. 6 b) of ref. \cite{FeOTal2016} is quite remarkable.}
\label{FANOBOTH}
\end{figure}

Fano factors calculated from CMD snapshots of the two CMD potentials are presented in Fig. \ref{FANOBOTH}. The dark grey areas delineate forbidden values of $F$ for magnetic atomic-sized contacts, while the areas underneath the light grey line are the forbidden values of $F$ for non-magnetic materials. Recall that $F$ is a measure of the number of partially open transmission channels in a contact, and the more channels contribute to the overall conductance, the more atoms are likely involved.

For the MEAM potential (Fig. \ref{FANOBOTH} b)), the calculated conductance values fall at or near the 4, 5 and 6 transmission channel lines. Experimentally, Fe has been shown to form last-contact structures with 6 transmission channels (see the experimental Fano diagram in Fig. 6 b) of ref. \cite{FeOTal2016}), which appears to indicate the formation of contacts with 3 or more atoms in them, according to the values of the Bratkovksy minimum cross-sections in table SI of the Supplemental Material \cite{SM}. The MEAM potential, with more covalent character, is thus seen to outperform the EAM potential in this regard, whose $F$ values for the 17 conductance values collected in table SII \cite{SM} are plotted in Fig. \ref{FANOBOTH} a).

Moreover, based on the low density of states of Fe at the Fermi level \cite{Garcia2000iron}, in comparison to Ni or Co, one would expect Fe to have a first maximum conductance peak at a lower conductance value than either of the latter metals. The experimental Fano diagram in Fig. 6 b) of ref. \cite{FeOTal2016} exhibits a significant number of conductance values at this expected low value of $\approx 1.2-1.4G_0$, but
in a histogram, they are subsumed by the broad peak at 
$\approx 2G_{0}$ (see fits to the histogram of conductance in the Supplemental Material \cite{SM}). Our interpretation, based on the simulation results presented above, is that slight differences in the structures responsible for this peak (see Fig. \ref{GeFEO}) could result also in deviations from the $2G_0$ value.

Therefore, we postulate that the discrepancy between the experiments and the simulations in the work of Vardimon \textit{et al.} \cite{FeOTal2016} for the case of Fe is the lack of CMD structures with minimum cross-sections above $\approx 3$ atoms at rupture when using an EAM potential, in a combination with a tight-binding model to calculate conductance. Using a more detailed model of the bonding between Fe atoms, such as the MEAM potential used in this work, should markedly improve the comparison between the experimental histogram of conductance and the one calculated in \cite{FeOTal2016}
%result in Fig. 3 c) of \cite{FeOTal2016}
based on the good agreement between the theoretical Fano diagram in Fig. \ref{FANOBOTH} b) and their experimental Fano diagram in Fig. 6 b) of  \cite{FeOTal2016}. 
 
It is also important to note that other BCC materials such as Ta, Mo and W, in similarity with iron, show experimental histograms of conductance with pronounced peaks at around $\approx 2G_0$ \cite{DenBoer2007,Halbritter2003}, and that the formation of similar structures could be explored in future work.
 
In summary, the pronounced peaks at $\approx 2G_{0}$ that appear in experimental histograms of conductance of body-centered cubic metal atomic-contacts could not be explained by considering single-atom contact structures \cite{ETESAMI2018}. Here, we show that DFT electronic transport calculations of structures with several atoms in the minimum cross-section give conductance values in agreement with experiments. Such structures arise when an energetically favorable reorientation occurs, from (001) to stable (011) layers perpendicular to the direction of stretching. We obtain this stable reorientation when using an interatomic potential (the MEAM potential) that includes directional bonding. Our findings contradict the presently-held assumption that the most likely stable pre-rupture contacts in BCC metals are made up of a single atom.  The MEAM potential thus provides a realistic mechanism of atomic rupture for Fe in which covalent bonding plays a key role. Body-centered cubic metals such as Fe may therefore represent good candidates for producing molecular junctions in which the electrode-molecule interface is an atomically flat surface.

\section{Acknowledgments}
This work was supported by the Generalitat Valenciana through PROMETEO2017/139 and GENT (CDEIGENT2018/028), the Spanish government through grants MAT2016-78625-C2-1-P,  FIS2016-80434-P and the Mar\'{i}a de Maeztu Programme for Units of Excellence in R\&D (MDM-2014-0377), by Comunidad Aut\'onoma de Madrid through Grant S2018/NMT-4321 (NanomagCOST-CM), by the Fundaci\'on Ram\'on Areces, and by the European Union Graphene Flagship under Grant No. 604391. JJP acknowledges the computer resources and assistance provided by the Centro de Computaci\'on Cient\' ifica of the Universidad Aut\'onoma de Madrid and the RES.

\bibliography{Iron}
\end{document}

\preprint{APS/123-QED}

\beginsupplement

\title{\vspace{-10cm}Directional bonding explains high conductance values of atomic contacts in bcc metals}

\author{W. Dednam}
\email{wd2@alu.ua.es}
\affiliation{Department of Physics, Science Campus, University of South Africa, Private Bag X6, Florida Park 1710, South Africa}
\affiliation{Departamento de F\'\i sica Aplicada and Unidad asociada CSIC, Universidad de Alicante, Campus de San Vicente del Raspeig, E-03690 Alicante, Spain}

\author{C. Sabater}
\email{carlos.sabater@ua.es}
\affiliation{Departamento de F\'\i sica Aplicada and Unidad asociada CSIC, Universidad de Alicante, Campus de San Vicente del Raspeig, E-03690 Alicante, Spain}

\author{M. R.Calvo}
\affiliation{Departamento de F\'\i sica Aplicada and Unidad asociada CSIC, Universidad de Alicante, Campus de San Vicente del Raspeig, E-03690 Alicante, Spain}

\author{C. Untiedt}
\affiliation{Departamento de F\'\i sica Aplicada and Unidad asociada CSIC, Universidad de Alicante, Campus de San Vicente del Raspeig, E-03690 Alicante, Spain}
 
\author{J. J. Palacios}
\affiliation{Departamento de F\'\i sica de la Materia Condensada, Condensed Matter Physics Center (IFIMAC), and
Instituto Nicol\'as Cabrera,  Universidad
Aut\'onoma de Madrid, 28049 Madrid, Spain} 
 
 \author{A. E. Botha}
\affiliation{Department of Physics, Science Campus, University of South Africa, Private Bag X6, Florida Park 1710, South Africa}
 
\author{M. J. Caturla}
 \affiliation{Departamento de F\'\i sica Aplicada and Unidad asociada CSIC, Universidad de Alicante, Campus de San Vicente del Raspeig, E-03690 Alicante, Spain}

\date{\today}% It is always \today, today,
             %  but any date may be explicitly specified   

\maketitle

\subsection{Introduction}

This supplemental material provides more details about the DFT calculations described in the main article, and also the results of conductance calculations on snapshots from CMD simulations performed with the MEAM and EAM potentials. Furthermore, it is shown how a sum of three Gaussian functions can be fitted to the experimental histogram in Fig. 1 c) of the main article. This shows that a variety of different stable structures contribute to the main peak in the experimental histogram.

\subsection{DFT calculations}

We extracted 33 representative snapshots of stable pre-rupture structures from the 100 CMD simulations with the MEAM potential, 10 of which correspond to the rupture process illustrated in Fig. 2 c) of the main article. The snapshots were trimmed down to $\sim$200 atoms centered on the minimum cross-section to allow conductance calculations to finish in a reasonable time. The results of the conductance calculations are shown in table \ref{TableMEAM}. Cases marked with an asterisk (*) in this table result in conductance values close to the first peak in the experimental histogram ($\approx 2G_{0}$) and correspond to structures such as those shown in Fig. 2 c) in the manuscript. On the other hand, 17 snapshots have been extracted from the 100 CMD rupture runs performed with the EAM potential, which were also trimmed down for conductance calculations. The results of these calculations are collected in table \ref{TableEAM}.

To improve the quality of conductance results, an all-electron basis set has been assigned to 15-20 atoms in the minimum cross-section of the trimmed-down inputs for conductance calculations. The all-electron basis set was optimized in \texttt{CRYSTAL14} \cite{CRYSTAL14} after adding uncontracted Gaussian-type orbitals to an existing basis set for Fe, and varying their coefficients and exponents in the same way as was done for Ni in ref. \cite{Doll2003}. The quality of the basis set has been verified by comparing the bandstructure it produces for bulk BCC Fe with that produced by \texttt{OpenMX} \cite{Ozaki2017}. 

\clearpage
\subsection{Tables }
\begin{table}[hbt!]
\begin{minipage}[t]{0.45\linewidth}
\caption{Contact type, Bratkvosky minimum cross-section and Conductance of snapshots from CMD simulations with MEAM potential for Fe. Cases marked with an asterisk (*) correspond to cross-sections with more than 2 atoms and conductance close to the experimental peak. }
\begin{tabular}[t]{ c c c c}
\toprule
\textrm{Rupture}& \textrm{Type} & \textrm{Min. cross-section} &\textrm{Conductance $(G_0)$}\\
\colrule
2 & 5-3-4 & 1.6 & 1.2 \\
5 & 9-7-8* &  4.7 & 2.4\\
8 & 8-6-9* & 3.4 & 2.3\\
11 & 2-1-2 & 0.6 & 1.0 \\
12 & 6-6-7* & 4.7 &	2.1\\
15 & 5-3-5 & 1.5 &	1.4\\
17 & 9-6-8* & 3.8 & 2.1\\
18 & 5-2-2-5 & 1.7 & 1.4\\
19 & 7-4-8* & 3.2 & 1.9\\
34 & 3-2-2-2 & 0.8 & 1.1\\
40 & 8-6-9* & 4.3 &2.1\\
46 & 4-2-4 & 1.5 & 1.5 \\
49 & 3-2-5 & 1.6 & 1.2 \\
51 & 4-2-4 & 1.6  & 1.3\\
54 & 6-3-6* & 3.0 & 2.3\\
55 & 4-2-2-5 & 1.2 & 1.3\\
56 & 5-2-4 & 1.5 & 1.3 \\
58 & 3-2-4 & 1.4 & 1.1 \\
60 & 3-2-4 & 1.6 & 1.6 \\
63 & 4-2-2-5 & 1.4 & 1.3 \\
64 & 5-2-2-5 & 1.7 & 1.9 \\
67 & 5-3-3-5 & 1.3 & 1.3 \\
70 & 3-1-3 & 0.8 & 1.2 \\
72 & 5-2-3 &  0.8 & 1.5 \\
73 & 5-2-3  & 1.4 & 1.1\\
74 & 3-1-3 & 0.9 & 1.0 \\
75 & 5-2-2-5 & 1.6 & 2.0 \\
77 & 8-7-8* & 4.0 & 2.3 \\
85 & 8-4-5* & 2.8 & 1.9\\
88 & 5-3-3-5 & 1.5 & 1.6 \\
91 & 7-6-10* & 4.2 & 2.5 \\
97 & 2-1-2 & 1.0 & 1.2 \\
99 & 4-2-3 & 1.2  & 1.7\\
\botrule
\label{TableMEAM}
\end{tabular}
\end{minipage}
\hfill
\begin{minipage}[t]{0.45\linewidth}
\caption{Contact type, Bratkvosky minimum cross-section and Conductance of snapshots from CMD simulations with EAM potential for Fe}
\begin{tabular}[t]{ c c c c}
\toprule
\textrm{Rupture}& \textrm{Type} & \textrm{Min. cross-section} &\textrm{Conductance $(G_0)$}\\
\colrule
1 & 4-3-4 & 2.5 & 1.2 \\
8 & 5-2-5 & 1.6 & 1.6 \\
10 & 3-2-2-3 &  1.5 & 1.0\\
13	& 5-3-3-5 & 1.7 & 1.4\\
16 	& 4-2-3  & 1.7 & 1.4	\\
21 & 3-2-3 & 1.5 & 1.4\\
23 & 3-2-3 & 1.7 & 1.5 \\
24 	& 4-3-6 & 1.6 & 2.0\\
42 	& 4-1-3 & 0.8 & 0.9\\
53 	& 2-1-3 & 0.7 & 1.0\\
69 & 3-1-2 & 0.7 & 0.8 \\
74 & 2-1-2 & 0.7 & 0.7 \\
77 & 3-1-2 & 0.7 & 0.8 \\
80 & 2-1-2 & 0.8 & 1.1 \\
87 & 2-1-2 & 0.6 & 0.7 \\
94 & 4-2-3 & 1.3 & 1.4 \\
98 & 2-1-2 & 0.5 & 0.7 \\
\botrule
\label{TableEAM}
\end{tabular}
\end{minipage}
\end{table}
\hfill

%\clearpage
\subsection{Gaussian Fit}
\begin{figure}[htp]
 \centering
 \includegraphics[width=0.5\textwidth]{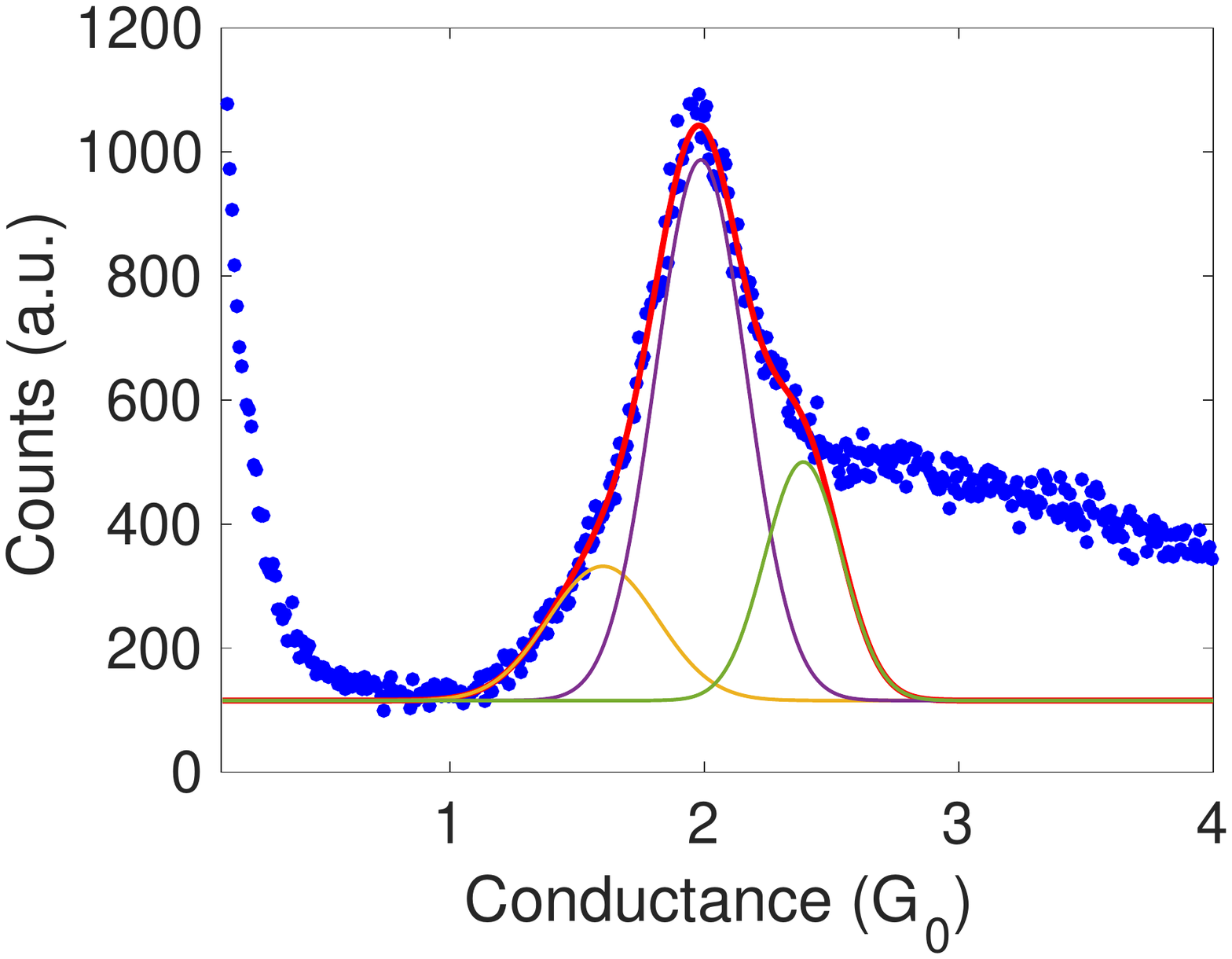}
\caption{A sum of three Gaussian functions fitted to the experimental conductance histogram obtained from the rupture of iron at 4.2 K. Blue markers are the raw experimental data, while the yellow, purple and green curves are Gaussian functions that sum to give the red line.}
\label{SMHis}
\end{figure}

Superimposed on the experimental histogram of conductance in Fig. 1 c) we have fitted three Gaussian curves -- see Fig. \ref{SMHis}. The blue markers represent the experimental data, and the red line through them,  the sum of the three Gaussian curves, which are  centered on 1.60, 1.99 and 2.39$G_0$, respectively, and plotted using golden, purple and green lines. 

The fact that we need more than 2 Gaussian curves in order to describe the main peak of the experimental data implies that there is not only a single and repeatable structure that gives rise to the experimentally observed peak. It also means that it is produced by different and a rich variety of structures. On the other hand the 3 centers of the Gaussians cover the range of calculated conductance values in table \ref{TableMEAM} rather well. The red line is expressed as

$$y(x)= \Phi+\sum_{i=1}^{3} a_{i}e^{\frac{x-b_{i}}{2c_{i}^2}},$$

where the coefficients $a_{i}$ are the amplitudes of the three underlying curve peaks, $b_{i}$ are the positions of their centers and $c_{i}$ are the standard deviations. The constant $\Phi$ is an offset with a value of 104 counts for this fit. Table \ref{Table:GAUFIT} collects all the fitting parameters.

\begin{table}[htp!]
\parbox{\linewidth}{
\caption{Fitting Parameters}
\begin{tabular}{ c c c c}
\toprule
\textrm{Gaussian}& \textrm{a} & \textrm{b} &\textrm{c}\\
\colrule
1 & 216 & 1.60 & 0.30  \\
2 & 872 & 1.99 & 0.25 \\
3 & 385 & 2.39 & 0.21 \\
\botrule
\label{Table:GAUFIT}
\end{tabular}}
\end{table}

\hfill

\bibliography{Iron}